\documentclass[aps,prl,reprint,amsmath,amssymb,superscriptaddress,floatfix]{revtex4-2}
\usepackage{graphicx}% Include figure files
\usepackage{hyperref}
\hypersetup{colorlinks, citecolor=blue, linkcolor=blue}
\usepackage{xcolor}
\usepackage{enumitem}
\usepackage{gensymb}
\newcommand{\KIEL}{Institute of Theoretical Physics and Astrophysics, University of Kiel, Leibnizstrasse 15, 24098 Kiel, Germany}
\newcommand{\HAM}{Department of Physics, University of Hamburg, 20355 Hamburg, Germany}

\begin{document}

%%%%%%%%%%%%%%%%%%%%%%%%%%%%%%%%%%%%%%%%
%       Header information
%%%%%%%%%%%%%%%%%%%%%%%%%%%%%%%%%%%%%%%%
\title{
Topological-chiral magnetic interactions in ultrathin films at surfaces
}

\author{Soumyajyoti Haldar}
\thanks{These two authors contributed equally}
\email[Email: ]{haldar@physik.uni-kiel.de}
\affiliation{\KIEL}

\author{Sebastian Meyer}
\thanks{These two authors contributed equally}
\altaffiliation[Current address: ]{Nanomat/Q-mat/CESAM, Universit{\'e} de Li{\`e}ge, B-4000 Sart Tilman, Belgium}
\affiliation{\KIEL}

\author{Andr\'e Kubetzka}
\affiliation{\HAM}

\author{Stefan Heinze}
\affiliation{\KIEL}

% It is always \today, today
\date{\today}

%%%%%%%%%%%%%%%%%%%%%%%%%%%%%%%%%%%%%%%%
%       Abstract
%%%%%%%%%%%%%%%%%%%%%%%%%%%%%%%%%%%%%%%%
\begin{abstract}
We demonstrate that topological-chiral magnetic interactions can play a key role
for magnetic ground states in ultrathin films at surfaces.
Based on density functional theory we show that significant chiral-chiral
interactions occur in hexagonal Mn monolayers due to large topological orbital
moments which interact with the emergent magnetic field. Due to the competition
with higher-order exchange interactions superposition states of spin spirals
such as the 2Q state or a distorted 3Q state arise.
Simulations of spin-polarized scanning tunneling microscopy images suggest
that the distorted 3Q state could be the magnetic ground state of a Mn monolayer 
on Re(0001).
\end{abstract}

\maketitle
%%%%%%%%%%%%%%%%%%%%%%%%%%%%%%%%%%%%%%%%%
% Introduction
%%%%%%%%%%%%%%%%%%%%%%%%%%%%%%%%%%%%%%%%%

Non-collinear spin structures are of fundamental interest in magnetism since 
they allow to obtain insight into the underlying microscopic 
interactions and are promising for spintronic applications \cite{Fert2017,Grollier2020}. For example, the
frustration of Heisenberg exchange interactions is the origin of spin spiral states found 
in many rare-earth elements or in the fcc-phase of Fe. The competition with 
beyond nearest-neighbor exchange occurs due to the long-range nature of the RKKY interaction characteristic for 
transition and rare-earth metals. Spin spirals can also arise 
due to the Dzyaloshinskii-Moriya (DM) interaction which occurs in materials with broken 
inversion symmetry such as surfaces or interfaces \cite{Bode2007}. 

Recently, more complex magnetic states such as the superposition of spin spirals 
\cite{Kurz2001,Hayami2017,Spethmann2020} or atomic scale spin lattices 
\cite{Heinze2011,Hoffmann2015,Bergmann2015} have raised much attention. They 
can occur due to terms beyond the pair-wise Heisenberg or DM interactions such 
as higher-order exchange interactions \cite{Takahashi77,MacDonald88,Hoffmann2020}.
They might also be driven by the recently proposed
topological-chiral \cite{Grytsiuk2020} and chiral multi-spin interactions 
\cite{Szunyogh2019,Brinker2019,Mankovky2020}. 
A prominent example of a multi-Q state -- a superposition of symmetry equivalent spin spirals (1Q states) -- is the 3Q state \cite{Kurz2001}. The 3Q state, which 
has first been predicted as the ground state
of a Mn monolayer on Cu(111) \cite{Kurz2001}, is an intriguing non-collinear 
spin structure on a two-dimensional lattice which leads to topological orbital moments 
and a topological Hall effect even in the absence of spin-orbit coupling \cite{Hanke2016}. 
It has also been predicted to trigger topological superconductivity in a conventional superconductor \cite{Bedow2020}. Recently, the
3Q state has been discovered in a Mn monolayer on the Re(0001) surface
by spin-polarized scanning tunneling microscopy (SP-STM) \cite{Spethmann2020}.
However, the locking of the spin structure to the atomic lattice could not be explained and 
it was speculated that a distortion of the 3Q state could be the origin.

Here we demonstrate based on density functional theory (DFT) that the recently
discovered topological-chiral magnetic interactions \cite{Grytsiuk2020} can be 
essential for the magnetic ground state of ultrathin films at surfaces. 
For freestanding Mn monolayers we show that large topological orbital moments occur
which interact with the emerging magnetic field leading to significant chiral-chiral
interactions. By the interplay of this term with higher-order exchange interactions complex spin structures such as the 2Q state or the distorted 3Q state can be stabilized. For Mn monolayers on Cu(111) and Re(0001) we predict similarly large
chiral-chiral interactions which could induce a distorted 3Q state 
magnetic ground state for Mn/Re(0001).
By simulating scanning tunneling microscopy images we 
show how the predicted magnetic ground states and thus the 
topological-chiral magnetic interactions may be verified experimentally.

We have used DFT as implemented in the {\tt FLEUR} code \cite{FLEUR} which
is based on the full-potential linearized augmented plane wave (FLAPW) method 
\cite{Weinert2009} to 
calculate total energies of non-collinear spin structures \cite{Kurz2004}
for unsupported Mn monolayers and on the Re(0001) surface.
The Mn monolayer has
been studied in hcp stacking on Re(0001) using asymmetric films with 6 layers of Re.
The films have been structurally relaxed as described in Ref.~\cite{Spethmann2020}. 
We have also used the {\tt VASP} code \cite{VASP2} which is based on the projector 
augmented wave (PAW) method \cite{VASP1,blo,blo1}. In these calculations we have 
also used asymmetric films for Mn/Re(0001) but have
increased the Re substrate to 10 layers. In addition, we have studied 
Mn/Cu(111) via {\tt VASP}. Calculations have been performed within
the local density approximation (LDA) and in the scalar-relativistic approximation,
i.e.~neglecting spin-orbit coupling. Further computational details can be found
in the supplemental material~\cite{supplement}.

\begin{figure}
\includegraphics[width=\linewidth]{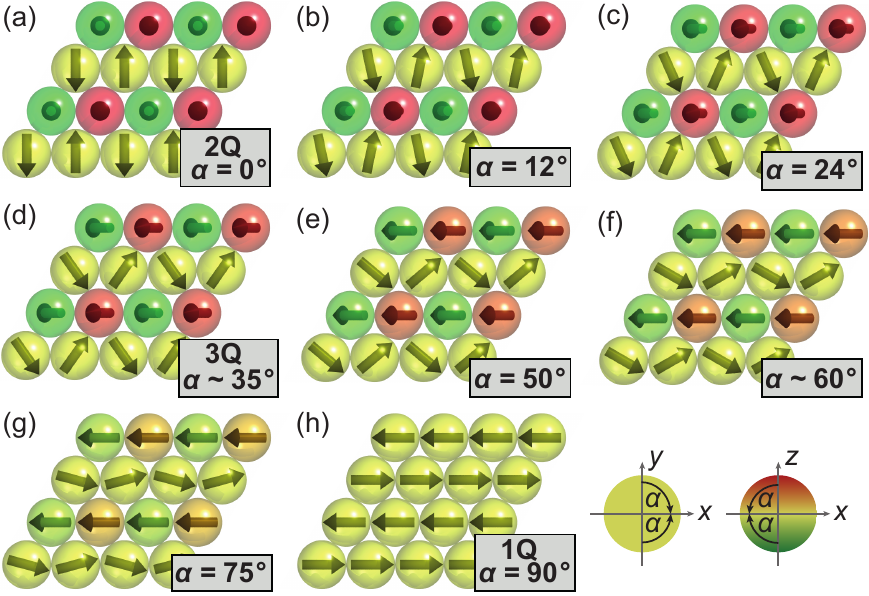}
\caption{Spin structures for a hexagonal Mn monolayer along the continuous path given 
by Eq.~(\ref{eq:path}) from (a) the 2Q state via (d) the 3Q to (h) the 1Q state.
%the 2Q-3Q-1Q path. 
Yellow arrows denote magnetic moments which rotate in the $xy$ plane (shown in the coordinate system above), red arrows denote moments in the positive $z$ direction, green arrows in the negative $z$ direction. The angle $\alpha$ is applied to rotate $\pm z$ spins into the $-x$ direction and $\pm y$ spins to the $+x$ direction. (a) 2Q state with $\alpha = 0^\circ$, (b) $\alpha = 12^\circ$ , (c) $\alpha = 24^\circ$, (d) 3Q state with $\alpha \sim 35^\circ$, (e) $\alpha = 50^\circ$, (f) $\alpha \sim 60^\circ$,
%which describes the energy minimum in Mn/Re(0001), 
(g) $\alpha = 75^\circ$, (h) 1Q state (RW-AFM)
%or row-wise antiferromagnetic state 
with $\alpha = 90^\circ$. }
\label{Figure: Spinstructures Conversion}
\end{figure}

%In order to cover a large part of the magnetic phase space 
For unsupported Mn monolayers we have calculated the energy dispersion of spin spirals since
they are the fundamental solution of the classical Heisenberg model. Spin spirals can 
be characterized by a vector ${\mathbf q}$ from the Brillouin zone (BZ). The 
calculated energy dispersion is qualitatively the same for all considered Mn
monolayers (see supplemental material~\cite{supplement}).
Among all spin spiral states the 
row-wise antiferromagnetic (RW-AFM) state (Fig.~\ref{eq:path}(h)) 
is the energetically lowest. 
The RW-AFM state can propagate along one
of the three symmetry equivalent directions of the surface corresponding to the
%. This correspondsto three energetically degenerate spin spiral states at the 
$\overline{\rm M}$
points of the two-dimensional BZ, i.e.~1Q states. One can construct linear combinations of 
two or three of these degenerate 1Q states under the constraint of fixed spin length 
at every lattice
site resulting in the 2Q (Fig.~\ref{Figure: Spinstructures Conversion}(a)) and the 3Q (Fig.~\ref{Figure: Spinstructures Conversion}(d)) 
state, respectively. These states are energetically degenerate with the 1Q states
within the classical Heisenberg model.

Figure \ref{Figure: Spinstructures Conversion} shows the spin structures which we consider 
in our total energy calculations to check for magnetic interactions beyond Heisenberg
exchange. We start from the 2Q state, Fig.~\ref{Figure: Spinstructures Conversion}(a), and rotate all spins in this structure continuously into the 1Q (RW-AFM) state, 
Fig.~\ref{Figure: Spinstructures Conversion}(h). This is achieved based on the
Rodriguez rotation formula such that the spin, $\mathbf{s}_i^\nu$, at site $i$ in the 
$\nu$-th rotation step is given by (see supplemental material for details~\cite{supplement}):
\begin{equation}
\mathbf{s}_i^\nu = \mathbf{s}_i^{\rm 2Q} \cos{\alpha_\nu} + \mathbf{s}_i^{\rm 1Q} \sin{\alpha_\nu},
\label{eq:path}
\end{equation}
where $\mathbf{s}_i^{\rm 2Q}$ and $\mathbf{s}_i^{\rm 1Q}$ are the spin directions
in the 2Q and 1Q state, respectively. The rotation angle $\alpha_\nu$ is 
varied from zero to $90^\circ$. The 3Q state corresponds to $\alpha_\nu \approx 35^\circ$,
Fig.~\ref{Figure: Spinstructures Conversion}(d).
It can be demonstrated that pair-wise Heisenberg exchange terms,
%\begin{equation}
% $H=-\sum\limits_{i,j} J_{ij} ( \mathbf{s}_i \cdot \mathbf{s}_j),$ 
 $-J_{ij} ( \mathbf{s}_i \cdot \mathbf{s}_j),$ 
%\end{equation}
where $J_{ij}$ are the exchange constants, do not vary along this path (see Supplemental Material~\cite{supplement}). Therefore, any change of the total energy obtained via DFT can 
%the observed change of the total DFT energies in Fig.~\ref{Figure: Mn UML} can 
only be explained by terms beyond the Heisenberg exchange. 

\begin{figure}[ht!]
\includegraphics[width=\linewidth]{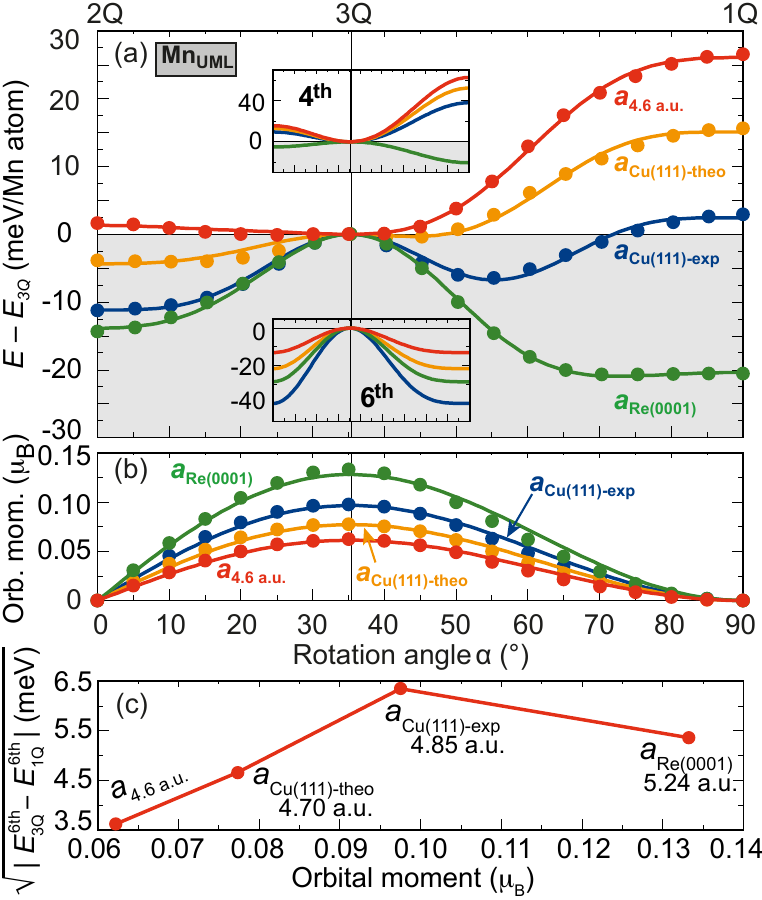}
\caption{(a) Energy along the 2Q-3Q-1Q path given by Eq.~(\ref{eq:path}) in hexagonal Mn unsupported monolayers (UMLs) for different in-plane lattice constants. Symbols denote total energies calculated from DFT using the {\tt FLEUR} code, the lines show the fit to higher-order-exchange interaction terms. 
The two insets show the contribution of the 4th and 6th order terms to the fit.
Red color shows the result for $a=4.6$~a.u., orange and blue color for the
theoretical and experimental in-plane lattice constant of Cu(111), respectively,
and green color the values for
the theoretical in-plane lattice constant of Re(0001) \cite{Ji2016}.
(b) Absolute value of the orbital moment per Mn atom along the direction perpendicular 
to the Mn UML. Symbols denote DFT values and lines the fit to the scalar spin chirality 
(see text for details).
(c) Orbital moment of the 3Q state vs.~the absolute value of the 
6th-order energy contributions at the 1Q w.r.t to the 3Q state 
(cf.~lower inset of Fig.~\ref{Figure: Mn UML}(a)).}
\label{Figure: Mn UML}
\end{figure}

Figure \ref{Figure: Mn UML}(a) shows the total DFT energies obtained for the path 
defined by Eq.~(\ref{eq:path}) via the {\tt FLEUR} code for a hexagonal unsupported monolayer 
(UML) of Mn with different in-plane lattice constants, i.e.~distance between nearest-neighbor
atoms. For the Mn UML on the lattice constant
of $a=4.6$~a.u., we find that the 3Q state is the energetically lowest spin configuration.
If we slightly increase the lattice constant to the value of the theoretical
Cu lattice constant, the 2Q state shifts below the 3Q state. In addition, a
small local energy minimum occurs close to the 3Q state. If the lattice constant is 
increased further to the experimental value of Cu the energy minimum of the distorted 3Q
state (Fig.~\ref{Figure: Spinstructures Conversion}(f)) becomes more pronounced
but the 2Q state remains the lowest state. For the even larger Re lattice constant
the energy of the 1Q state has moved below the 2Q state. There is a tiny energy
minimum for a distorted 1Q state with an angle of $\alpha \approx 75^\circ$ 
(Fig.~\ref{Figure: Spinstructures Conversion}(g)).

The higher-order exchange interactions (HOI) of fourth order are the biquadratic interaction
and the three-site and four-site four spin interactions \cite{Hoffmann2020}:
\begin{widetext}
\vspace*{-\baselineskip}
\begin{eqnarray}
      H_{\rm 4th}  = &  - & \sum\limits_{i,j} B_{ij} ( \mathbf{s}_i \cdot \mathbf{s}_j)^2 
    - \sum_{ijk} Y_{ijk} [(\mathbf{s}_i \cdot \mathbf{s}_j)(\mathbf{s}_j \cdot \mathbf{s}_k) + (\mathbf{s}_j \cdot \mathbf{s}_i)(\mathbf{s}_i \cdot \mathbf{s}_k)+
    (\mathbf{s}_i \cdot \mathbf{s}_k)(\mathbf{s}_k \cdot \mathbf{s}_j)] + \nonumber \\
    & - &  \sum_{ijkl}  K_{ijkl} [(\mathbf{s}_i \cdot \mathbf{s}_j)(\mathbf{s}_k \cdot \mathbf{s}_l) + %\nonumber \\
    %& - & \sum_{ijkl}  K_{ijkl}[(\mathbf{s}_i \cdot \mathbf{s}_j)(\mathbf{s}_k \cdot \mathbf{s}_l)+
  %& + &  
  (\mathbf{s}_i \cdot \mathbf{s}_l)(\mathbf{s}_j \cdot \mathbf{s}_k) %\nonumber \\
   -   (\mathbf{s}_i \cdot \mathbf{s}_k)(\mathbf{s}_j \cdot \mathbf{s}_l)],
 \label{eq:HOI}
\end{eqnarray}
\vspace*{-\baselineskip}
\end{widetext}
where $B_{ij}, Y_{ijk}$ and $K_{ijkl}$ are the corresponding higher-order exchange constants.

The energy of these three terms varies along the given path according to the function 
$E_{\rm 4th}(\alpha)=\kappa_{\rm 4th} (2 \cos^4 {\alpha} + 6 \sin^4{\alpha}-4 \cos^2 {\alpha} \sin^2 {\alpha})$ with a strength $\kappa_{\rm 4th}$
(see Supplemental Material~\cite{supplement}). This function
has local extrema at the 1Q, 2Q, and 3Q state (see upper inset of 
Fig.~\ref{Figure: Mn UML}(a)). 
Since the 2Q state is always in between the 1Q and the 3Q state, 
i.e.~it cannot be the lowest state as long as higher-order terms are restricted
to fourth order, it has been excluded in previous DFT
calculations for Mn UMLs and Mn/Cu(111) \cite{Kurz2001}. However, 
the total energy calculations for Mn UML (Fig.~\ref{Figure: Mn UML}(a)) 
cannot be explained based on the higher-order interactions given by Eq.~(\ref{eq:HOI}). 
Note that the DFT calculations were performed in the scalar-relativistic approximation such 
that chiral-multi spin interactions \cite{Szunyogh2019,Brinker2019,Mankovky2020} cannot occur.

Recently, it has been demonstrated that due to the interaction of the topological orbital moment,
which can arise in non-collinear spin structures even in the absence of spin-orbit coupling \cite{Hoffmann2015,Hanke2016},
with the emergent magnetic field the so-called chiral-chiral interaction occurs \cite{Grytsiuk2020}:
\begin{equation}
    H_{\rm CC} = - \sum\limits_{ijk} \kappa^{\rm CC}_{ijk} 
    [\mathbf{s}_i \cdot (\mathbf{s}_j \times \mathbf{s}_k)]^2
\end{equation}
with a site-dependent interaction strength $\kappa^{\rm CC}_{ijk}$.
This 6th order interaction scales with the square of the scalar spin chirality
$\chi_{ijk}=\mathbf{s}_i \cdot (\mathbf{s}_j \times \mathbf{s}_k)$ which varies for
the given path at every site as $\chi_{ijk} \propto \cos^2 {\alpha} \sin{\alpha}$ 
(see Supplemental Material~\cite{supplement}). Therefore, the chiral-chiral interaction is 
proportional to $\cos^4{\alpha} \sin^2{\alpha}$ (lower inset of Fig.~\ref{Figure: Mn UML}(a)). 

In order to check the importance of this interaction in our Mn UMLs we have calculated via 
DFT the topological orbital moment, which is proportional to the scalar spin chirality 
(Fig.~\ref{Figure: Mn UML}(b)). As expected
it exhibits a maximum at the 3Q state and vanishing values for the 2Q and 1Q state and 
is oriented perpendicular to the film. The orbital moment for the 3Q state is in good
agreement with previous calculations \cite{Hanke2016}. An excellent fit is achieved 
for the angle dependent variation of the orbital moment  (Fig.~\ref{Figure: Mn UML}(b))
by using the analytical form of the scalar spin chirality given above.

The competition of 4th- and 6th-order interactions
%the higher-order and the chiral-chiral interaction 
can explain 
the observed trend of the DFT energy curves for the Mn UMLs as shown by the insets
of Fig.~\ref{Figure: Mn UML}(a). The 4th-order contribution favors the 3Q
state for the three lower lattice constants and the 1Q state for the Re lattice
constant. The 6th-order interaction, on the other 
hand, favors the 1Q and the 2Q state and drives the transition to the distorted 3Q state. 

The trend of DFT curves in Fig.~\ref{Figure: Mn UML}(a) is captured by considering 
only the 4th- and 6th-order term (see supplemental material~\cite{supplement}).
To improve the fit,
%For a perfect fit of the DFT energycurves shown in Fig.~\ref{Figure: Mn UML}(a)
we added contributions from 8th- and 10th-order terms in 
Fig.~\ref{Figure: Mn UML}(a). Such an expansion of a spin model for itinerant
magnets in a power series of cosines between spins has been proposed 
\cite{Rosengaard1997} starting from the Liechtenstein formula \cite{Liechtenstein1987}.

To show the significance of the chiral-chiral interaction, we take a closer look to the 6th-order term 
(lower inset of Fig.~\ref{Figure: Mn UML}(a)). The energy difference between the 3Q and the 1Q state 
of the 6th-order term increases proportional to the square of the topological orbital moment from 
$a=4.6$~a.u.~up to $a=4.85$~a.u.~(Fig.~\ref{Figure: Mn UML}(c)).
This is expected since the chiral-chiral interaction is proportional to the square of the scalar
spin chirality while the orbital moment is linear with this quantity. 
%The strength of the 6th-order term (Fig.~\ref{Figure: Mn UML}(c)) increases proportional 
%to the square of 
%As expected its strength increases proportional  to the square of the increase 
%the topological orbital moment from $a=4.6$~a.u.~up to $a=4.85$~a.u.~as 
%expected for the chiral-chiral interaction.
At the Re lattice constant, $a=5.24$~a.u., the orbital moment
further increases, while the energy contribution from the 6th order term
decreases. Due to the large increase of the lattice constant the local 
density of states changes significantly (see supplemental material~\cite{supplement}) and
thereby $\kappa_{ijk}^{\rm CC}$, which depends on the electronic structure.

Note, that the bicubic interaction, $( \mathbf{s}_i \cdot \mathbf{s}_j)^3$, which
is possible in systems with a spin moment of $\ge 3 \mu_B$ \cite{Hoffmann2020}, 
could also contribute to the 6th order energy term
since it exhibits the same functional form along the given path (see supplemental material~\cite{supplement}). 
However, the large topological orbital moments and the
scaling of the 6th-order energy with the orbital moment
provide strong evidence that the chiral-chiral interaction is dominating.

\begin{figure}[hbt!]
\includegraphics[width=\linewidth]{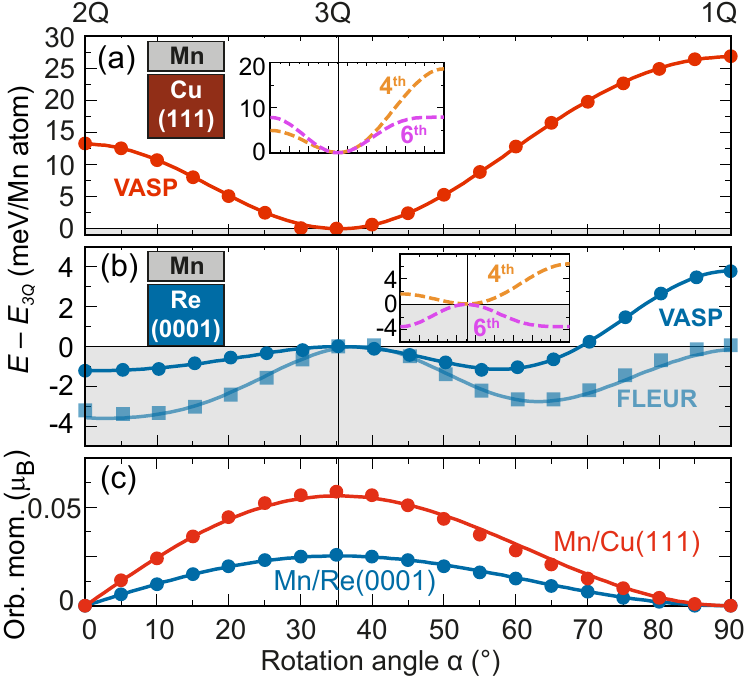}
\caption{Energy along the continuous 2Q-3Q-1Q path given by Eq.~(\ref{eq:path}) in
(a) Mn/Cu(111) and (b) Mn/Re(0001). Filled circles are energies calculated from DFT, the lines show the fit to higher-order-exchange interaction (insets show 4th and
6th order terms). 
For Mn/Cu(111) DFT calculations were performed using the {\tt VASP} code and for Mn/Re(0001) both the {\tt FLEUR} (squares) and the {\tt VASP} code (circles) were applied.
%For Mn/Cu(111) DFT calculations were performed using the {\tt VASP} code and for Mn/Re(0001) both methods, i.e.~the FLAPW method implemented in the {\tt FLEUR} code (squares) and the PAW method implemented in {\tt VASP} (circles) were applied.
(c) Absolute value of the orbital moment per Mn atom directed perpendicular to the surface. Symbols denote DFT values and lines the fit to the scalar spin chirality.}
\label{Figure: MnRe(0001)}
\end{figure}
Now we turn to Mn monolayers on surfaces. The 3Q state has first been predicted
for Mn/Cu(111) \cite{Kurz2001}, however, the 2Q state had not been considered
in that study. The DFT total energy curve along the path of Eq.~(\ref{eq:path})
(Fig.~\ref{Figure: MnRe(0001)}(a)), obtained via
{\tt VASP}, is consistent with the expectation of Ref.~\cite{Kurz2001}, i.e.~the
3Q state is lowest and the 2Q state lies between the 3Q and the 1Q state. 
However, a good fit of this energy curve requires to take not only the 4th
order exchange interactions into account but also the 6th order terms
(inset of Fig.~\ref{Figure: MnRe(0001)}(a)). The topological orbital
moments are of the same order of magnitude as for the UMLs and can be fit by the 
scalar spin chirality (Fig.~\ref{Figure: MnRe(0001)}(c)). 
This shows that the chiral-chiral interaction is significant in this system.

Experimentally, the 3Q state was discovered in hcp-Mn/Re(0001) \cite{Spethmann2020}.
From previous DFT calculations it is known that the RW-AFM state (1Q state) is energetically lowest
among all spin spiral states \cite{Spethmann2020}. The 3Q state formed from a 
superposition of the three equivalent RW-AFM states \cite{Kurz2001} 
is even slightly lower in total energy. 
%We have calculated the total energy for this
%system along the path defined by Eq.~(\ref{eq:path}). 
Surprisingly, the perfect 3Q state is only a local energy maximum along the path
given by Eq.~(\ref{eq:path})
while there are two local energy minima: one at the 2Q state and one for a distorted
3Q state (Fig.~\ref{Figure: MnRe(0001)}(b)). 
In the {\tt FLEUR} calculation the 2Q state is slightly lower in energy
and in {\tt VASP}
%which has been performed for a Re substrate of 10 layers 
the two states are energetically nearly degenerate. 

Similar to the case of the UMLs a good fit of the energy curve of Mn/Re(0001)
can only be obtained if higher-order terms beyond fourth order
%, i.e.~those given in Eq.~(\ref{eq:HOI}), 
are taken into account (inset of Fig.~\ref{Figure: MnRe(0001)}(b). 
In particular, the 6th order terms, corresponding to the topological-chiral or bicubic interaction, are decisive to capture the two local energy minima. The topological
orbital moments are smaller for Mn/Re(0001) (Fig.~\ref{Figure: MnRe(0001)}(c))
than for Mn/Cu(111) and so is the 6th order energy contribution. 
%Interestingly, the sign of the 6th order term is opposite between the two systems
%which could be due to a change of the sign of the interaction strength or due to
%the competition of bicubic and chiral-chiral interactions.
%Fig.~\ref{Figure: MnRe(0001)} 
%shows that the effect reported for UMLs is also robust for Mn monolayers on surfaces. 

\begin{figure}[htbp!]
\includegraphics[scale=1]{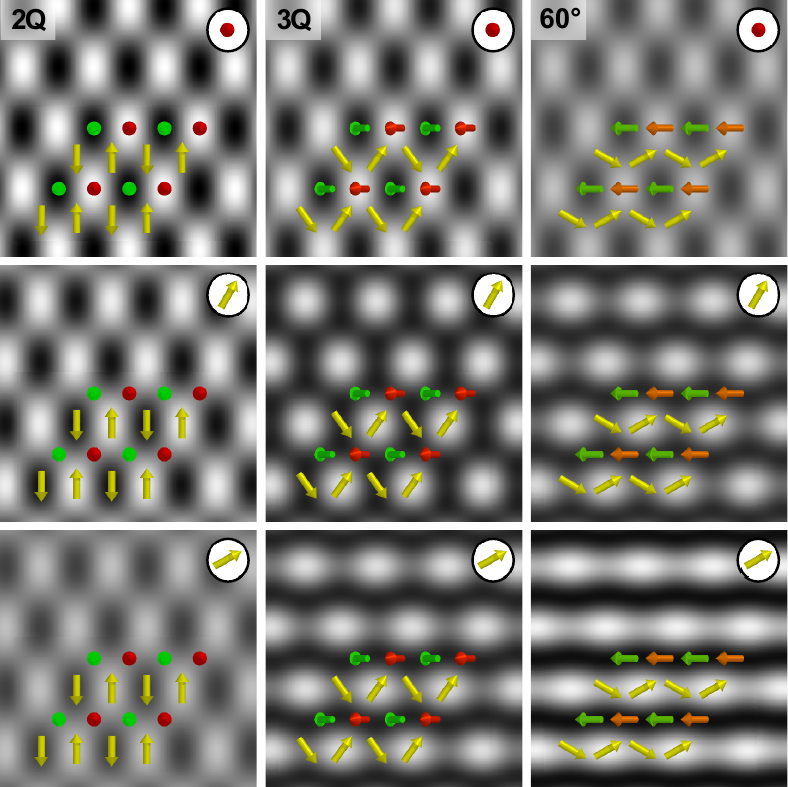}
\caption{Simulated spin-polarized scanning tunneling microscopy images at constant height 
(0.8 nm) for the 2Q, the 3Q, and the canted 3Q (60\degree) spin state for 3 different tip magnetization directions. The simulations have been performed using the model described
in Ref.~\cite{Heinze2006}. 
%Note that SP-STM images obtained directly from our DFT
%calculations are qualitatively the same (see Supplemental Material~\cite{supplement}).
All panels have the same color scale of 1.5 pm from black to white.}
\label{Figure:STM}
\end{figure}

Complex non-collinear spin structures in ultrathin films as proposed 
in this work can be resolved
down to the atomic scale using SP-STM \cite{Wortmann2001,Heinze2006,Bergmann2014}.
In order to check whether the three magnetic states compared above can be
distinguished in an experiment, we have simulated SP-STM images for the
2Q state, the 3Q state, and the distorted 3Q state for three different 
magnetization directions of the STM tip (Fig.~\ref{Figure:STM}). For the
2Q state the contrast is very similar for the three cases which indicates
that different rotational domains will look the same. For the 3Q state,
in contrast, the SP-STM image for an out-of-plane magnetized tip 
(middle panel in the upper row of Fig.~\ref{Figure:STM}) is qualitatively
different from that for a tip with an in-plane magnetization component
(lower two panels in the middle row). Thereby, the 2Q and the 3Q state
can be clearly distinguished if different rotational domains are imaged
or if the tip magnetization is rotated by an external magnetic field.

In contrast, the SP-STM images of the distorted 3Q state are very similar
to those of the ideal 3Q state and much harder to distinguish experimentally.
However, the perfect 3Q state exhibits only a very weak coupling to the atomic
lattice and the energetically preferred rotation of the 3Q state is in contrast 
to SP-STM experiments for Mn/Re(0001) \cite{Spethmann2020}. Therefore, it has
been proposed that a distorted 3Q state -- such as the one found here -- 
may explain the experimental observation
since its reduced symmetry enhances the coupling to the atomic lattice.

In conclusion, we have demonstrated that the competition between higher-order
exchange interactions and topological-chiral interactions needs to be taken into 
account in Mn monolayers. The novel types of magnetic ground states such as the 
%can lead to novel types of magnetic ground states in Mn monolayers.
2Q or the distorted 3Q state will affect the
transport properties of such systems \cite{Hanke2016} 
as well as their coupling to adjacent superconductors \cite{Bedow2020}.
We anticipate that topological-chiral magnetic interactions play an
important role for other film systems.
%Our work shows that previously
%applied spin models of itinerant magnets have neglected important terms.

We gratefully acknowledge computing time at the supercomputer of the North-German Supercomputing Alliance (HLRN) and financial support from the Deutsche 
Forschungsgemeinschaft (DFG) via Project No.~418425860 and No.~408119516.
It is our pleasure to thank Yuriy Mokrousov, Kirsten von Bergmann,
Lydia St\"uhmer-Herrmann, Moritz Goerzen, Mara Gutzeit, and Souvik Paul 
for valuable discussions.

%%%%%%%%%%%%%%%%%%%%%%%%%%%%%%%%%%%%%%%%
%   References
%%%%%%%%%%%%%%%%%%%%%%%%%%%%%%%%%%%%%%%%
\bibliography{Literature}

\end{document}